\documentclass[doublecol]{ctextemp_epl2}
\usepackage{mathrsfs}
\usepackage{amssymb}
\usepackage{amsfonts}
\usepackage{amsmath}
\usepackage{latexsym}
\usepackage{graphicx}
\usepackage{amsfonts,amssymb}

\title{Exact Solution for Optimal Navigation with Total Cost Restriction}
\shorttitle{Exact Solution for Optimal Navigation with Total Cost Restriction} %Insert here a short version of the title if it exceeds 70 characters

\author{Y. Li, D. Zhou, Y. Hu\footnote{E-mail: yanqing.hu.sc@gmail.com}, J. Zhang \and Z. Di}
\shortauthor{Y. Li\etal}

\institute{Department of Systems Science, School of Management, Beijing Normal University - Beijing 100875, PRC\\
} 

\abstract{Recently, Li \textit{et al.} have concentrated on
Kleinberg's navigation model with a certain total length constraint
$\Lambda  = cN$, where $N$ is the number of total nodes and $c$ is a
constant. Their simulation results for the 1- and
2-dimensional cases indicate that the optimal choice for adding
extra long-range connections between any two sites seems
to be $\alpha=d+1$, where $d$ is the dimension of the lattice and
$\alpha$ is the power-law exponent. In this paper, we
prove analytically that for the 1-dimensional large networks, the
optimal power-law exponent is  $\alpha=2$ Further, we study the
impact of the network size and provide exact solutions
for time cost as a function of the power-law exponent $\alpha$. We
also show that our analytical results are in excellent agreement
with simulations.}

\begin{document}
\maketitle

\section{Introduction}
Since Milgram and his cooperators conducted the first
small-world experiment in the 1960s, much attention has
been dedicated to the problem of navigation in real social networks.
The first navigation model was proposed by Kleinberg\cite{Kle00}. He
employed an $L \times L $  square lattice, where in addition to the
links between nearest neighbors each node $i$ was
connected to a random node $j$ with a probability $P_{ij}\propto
r_{ij}^{-\alpha}$ ($r_{ij}$ denotes the lattice distance
between nodes $i$ and $j$). Kleinberg has
proved that when $\alpha=d$, where $d$ is the dimension of the
lattice, the optimal time cost of navigation with a decentralized
algorithm is at most $O(\log^2 L)$. The optimal case
indicates that individuals are able to find short paths
effectively with only local information, which can explain six
degrees of separation quite well. In recent years, further studies
on Kleinberg's navigation model have been developed \cite{Rob06,
Car09A, Hav10, Car09B, Bar01,Yang10,entropy}. Roberson \textit{et
al.} study the navigation problem in fractal small-world networks
\cite{Rob06}, where they prove that $\alpha=d$ is also the
optimal power-law exponent in the fractal case. Cartozo \textit{et
al.} use dynamical equations to study the process of Kleinberg
navigation \cite{Car09A, Car09B}. They provide an exact solution for
the asymptotic behavior of such a greedy algorithm as a function of
the dimension $d$ of the lattice and the power-law exponent
$\alpha$. Yang \textit{et al}. construct a network with limited
cost $\Lambda = C$. The limited cost $\Lambda$
represents the total length of the long-range connections which are
added with power-law distance distribution $P(r) = a{r^{ - \alpha}}$
\cite{Yang10} in one-dimensional space. They find that the network
has the smallest average shortest path when $\alpha = 2$. More
recently, Li \textit{et al}. have considered Kleinberg's navigation
model with a total cost constraint \cite{Hav10}. In their model, the
total length of the long-range connections is restricted
to $\Lambda = c \cdot N$, where $N$ represents the total number of
nodes in the lattice based network and $c$ is a positive constant.
Their results show that the best transport condition(minimal number
of steps to reach the target) is obtained with a
power-law exponent $\alpha =d+1$ for constrained navigation in a
$d$-dimensional lattice in the 1 and 2-dimensional cases.
In this paper, we give a rigorous theoretical analysis of
the optimal condition $\alpha=2$ for navigation and provide the
exact time cost for various power-law exponents $\alpha$ on the
1-dimensional cost constrained network.

\section{Dynamical Equations for One-Dimensional Navigation with Cost Restriction}

\begin{figure}
\center
\includegraphics[width=5cm]{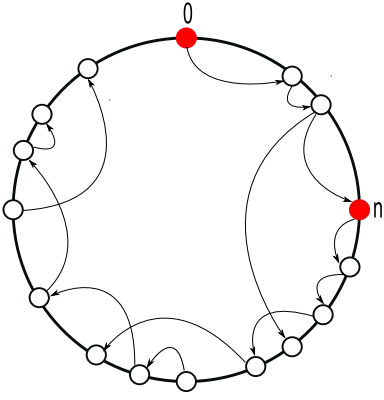}
\caption{The navigation model in this paper. Node $0$ is the
starting node, and node $n$ is the target.}\label{network}
\end{figure}

We consider the one-dimensional navigation problem on a cycle with
$N=4n$ nodes. For the sake of simplicity, we assume that each node
has only two short-range connections to its two nearest neighbors,
and the probability of having a long-range connection satisfies to a
certain power-law distribution. Obviously, the largest possible
length of a long-range connection is $2n$ in this cycle. We number
all nodes inclusively from $0$ to $4n-1$ and assume that the
navigation process starts from the node $0$ and ends at the node $n$
for further simplification. The network is illustrated in
FIG.\ref{network}.

According to the discussion above, the
probability of a long-range connection between any given pair of nodes with a distance $r$ is
\begin{equation}\label{prob of K}
p(r,\alpha) = \frac{{r^{ - \alpha} }}{2{\sum\limits_{r = 1}^{2n} {r^{ - \alpha} }
}}, \alpha \ge 0,
\end{equation}
where $\alpha$ is the power exponent of the power-law distribution. The expected length of the long-range connection from
any node satisfies $E(L_\alpha)=2\sum\limits_{r = 1}^{2n} {r \cdot
p(r,\alpha)}$. To be consistent with Li's work\cite{Hav10}, in this
paper we set the total cost limit to $\Lambda = c \cdot 4n$, where
$c$ is a positive constant and $4n$ is the number of nodes on the
cycle. Subject to this limit, the expected number of long-range
connections on the whole cycle can be written as $E(N_\alpha) {\rm{ =
}}\frac{\Lambda }{{E(L_\alpha )}}$.

Since all nodes are homogeneous, we know that the number of directed
long-range connections from each node should obey the Poisson
distribution with a parameter $\lambda=\frac{{E(N_\alpha)}}{4n}$.
Thus, the probability of a one long-range connection from
each node can be given by $\lambda e^{- \lambda} $. When $\lambda$
is small enough, the probability of the existence two or more than two
long-range connections for a arbitrary node can be ignored. More
over, for a given node and distance $r$, there are only two nodes
which satisfy the condition that the distances between them and the given node be
$r$. So, if $\lambda$ is too large, we cannot construct a spatial
network on which the length of long-rang connections is power law
distribution. According to the above two reasons, in this paper we
only consider the case where navigation process is carried out by at
most one long-range connection ($c$ is small) for each node.

If we use $E(L_s^\alpha)$ to denote the expected distance by a long-range
connection from a node $s$ toward the target node $n$, then we have
\begin{equation}\label{long-range length of s}
E(L_s^\alpha ) = \sum\limits_{r = 1}^{n - s} {r \cdot p(r,\alpha)}  +
\sum\limits_{r = n - s + 1}^{2n - 2s - 1} {(2n - 2s - r) \cdot
p(r,\alpha)}.
\end{equation}
We further denote the expected distance by an edge (long or short
range) from a node $s$ toward the target node $n$ as $E(J_s)$. Then
we have
\begin{equation}\label{Ejs}
E(J_s ) = \lambda e^{- \lambda} \cdot E(L_s^\alpha ) + 1 - \lambda e^{-
\lambda} \cdot [\sum\limits_{r = 1}^{n - s} {p(r,\alpha)}  +
\sum\limits_{r = n - s + 1}^{2n - 2s - 1} {p(r,\alpha)} ]
\end{equation}

For simplicity, we consider the continuous form of all equations provided above. Then $\lambda$ can
be written as
\begin{equation}\label{lambda}
 \lambda  = \begin{cases}
 c\frac{2}{{2n + 1}},&\alpha = 0 \\
 c\frac{{\ln 2n}}{{2n}},&\alpha = 1 \\
 c\frac{{1 - \frac{1}{{2n}}}}{{\ln 2n}},&\alpha = 2 \\
 c\frac{{\alpha - 2}}{{\alpha - 1}} \cdot \frac{{{{(2n)}^{1 - \alpha}} - 1}}{{{{(2n)}^{2 - \alpha}} - 1}},&else. \\
\end{cases}
\end{equation}
When the network size is large enough, Eq.(\ref{lambda}) can be simplified to
\begin{equation}\label{lambda2}
 \lambda  = \begin{cases}
 0,&0 \le \alpha \le 2 \\
 c\frac{{\alpha - 2}}{{\alpha - 1}},&else. \\
\end{cases}
\end{equation}

The method of dynamical equations is used to deduce the searching time with limited total cost. Suppose that at time $t$, the
corresponding position is $s(t)$. Obviously, $s(0)=0$ holds. The
dynamical equation can be written as
\begin{equation}\label{time complexity}
\left\{ \begin{array}{l}
\frac{{ds}}{{dt}} = E(J_s ), \\
s(0) = 0. \\
\end{array} \right.
\end{equation}

Before solving Eq.(\ref{time complexity}), we first study
the optimal power-law exponent $\alpha$ by comparing $E(J_s)$
(Eq.(\ref{Ejs})) under different values of $\alpha$. We rewrite the
distance to the target $n-s$ as $\varepsilon n$, where
$0<\varepsilon \le1$ is a constant. For any given $0<\varepsilon
\le1$, we assume $\varepsilon n$ is large enough, such
that long-range connections will be used in the search process.
Finally, Eq.(\ref{Ejs}) can be simplified to the following
forms,

\begin{equation}\label{simplification of Ejs}
 E(J_s ) \sim \begin{cases}
 \frac{1}{4}\varepsilon^2 c + 1,&\alpha = 0 \\
 \frac{{\ln 2}}{2}\varepsilon c + 1,&\alpha = 1 \\
 \frac{1}{2}c + 1,&\alpha = 2 \\
 \frac{e^{- c \frac {{\alpha-2}}{{\alpha-1}}} }{{2(\alpha - 1)}}  c + 1,&\alpha>2 \\
 \frac{{2^{\alpha - 1}  - 1}}{{2(\alpha - 1)}}\varepsilon^{2 - \alpha} c + 1,&else \\
\end{cases}
\end{equation}

It is not difficult to show the right side of the
Eq.(\ref{simplification of Ejs}) monotonically increases with $\alpha$ for $0\le \alpha\le 2$ and decreases with $\alpha$ for $\alpha>2$. We can also prove that $E(J_s)$
is continuous at the point $\alpha=2$. Overall, $E(J_s)$ is
continuous and  reaches its maximal value at $\alpha=2$ for any given
$\varepsilon$. It has been revealed that the optimal condition for navigation with limited cost is a tradeoff between the length and the number of long-range connections added to the cycle. Prior to solving the dynamical equations theoretically, we have already shown that the optimal power-law exponent is $\alpha=2$ with some proper simplifications. 

\begin{figure}
\center
\includegraphics[width=7cm]{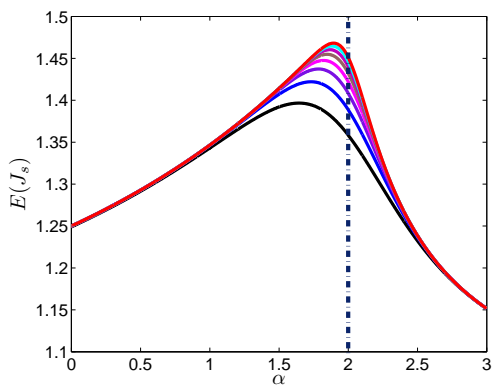}
\caption{Effects of the network size $n$ on the numerical
results. The network size is $10^{\beta}$, and $\beta$ is chosen
from 3 to 10 from bottom to top. It can be seen that the optimal
choice of $\alpha$ gets closer to $\alpha=2$ as $n$
increases.}\label{varn}
\end{figure}

\section{Results}

\begin{figure}
\center
\includegraphics[width=7cm]{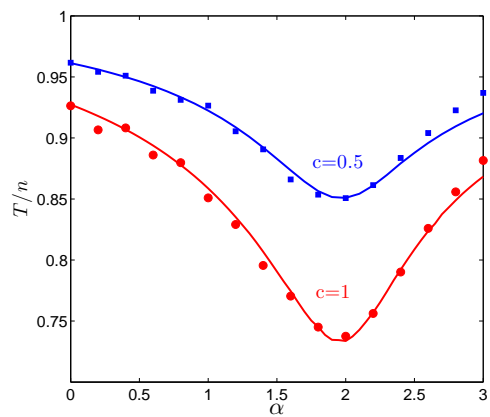}
\caption{Comparison between numerical results and simulation results
for $c=0.5$ and $c=1$ respectively. The curves represent the
numerical results while the squares denote the simulation results
with $n=10000$. It is shown that our numerical solutions are
consistent with the simulation results and both of them have the
optimal value at $\alpha=2$ approximately.}\label{compared}
\end{figure}

As discussed above, the
optimal power-law exponent for navigation in a 1-dimensional large network is
$\alpha=2$. FIG.\ref{varn} represents the size effect
on $E(J_s)$. It can be found that
$\alpha=2$ is the optimal power-law exponent when
the network size goes to infinity.

To obtain exact time cost of navigation with cost restriction,
 Eq.(\ref{time complexity}) should be solved. In the following, we
will first give its numerical results and then derive its exact
solutions for various values of $\alpha$. The Ronge-Kutta
method has been introduced to solve the dynamical equation
numerically and the results are presented in FIG.\ref{compared}. It
shows that the optimal power-law exponent is $\alpha=2$. To check up
our method, we also perform search experiments on the
one-dimensional cycle. The comparison between our analytical results
and the simulation results with $c=0.5$ and $c=1$ are given by
FIG.\ref{compared}. As can be seen, they agree quite well and both
of them obtain the optimal navigation at $\alpha=2$.

It is able to get the exact solutions of navigation with limited cost for various values of $\alpha$. For instance, the dynamical equation in the case $\alpha=0$ is,
\begin{equation}\label{r=0}
\frac{{ds}}{{dt}} = \frac{{c(n - s - 1)^2 }}{{4n^2}} + 1.
\end{equation}
Based on the initial condition $s(0)=0$, we can get the exact solution of Eq.(\ref{r=0}).

Here, we use $T$ to denote the time cost for getting the destination
node $n$. We should have $T = \frac{2n}{{\sqrt c }}\arctan \frac{{\sqrt c
}}{2}$ for large enough $n$. Thus, the average required time to navigate from the source
node to the target satisfies
\begin{equation}\label{T_r=0}
\frac{T}{n} = \frac{2}{{\sqrt c }}\arctan \frac{{\sqrt c }}{2}.
\end{equation}

\begin{figure}
\center
\includegraphics[width=7cm]{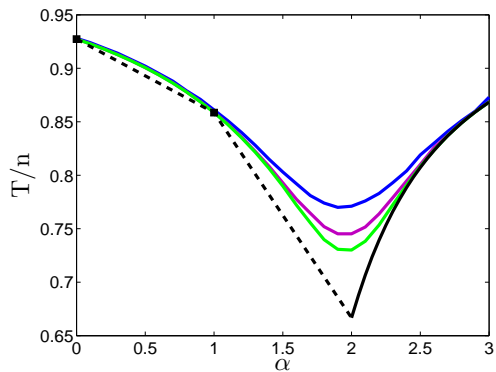}
\caption{Size effect on navigation. The exact solutions are
 obtained on the 1-dimensional cycle with parameter $c=1$. The network
  size $n$ is $10^{\beta}$, and parameter $\beta$ values of the upper
  three solid curves are chosen from 3 to 5 from top to bottom.
  It is shown that $\alpha=2$ is always the optimal choice for
  various values of $n$. The bottom line, obtained when $n$ goes
  infinity, represents the exact solutions of the navigation process
  with limited cost. Notice that we only provide the exact solutions
  at $\alpha=0$ and $\alpha=1$ when $\alpha<2$, thus we only connect
  the two exact points with dashed lines. }\label{sizeeffect}
\end{figure}

Analogously, the exact solution for exponent $\alpha$ when
$n$ approaches infinity are acquired as
\begin{equation}\label{time cost}
 \frac{T}{n} = \begin{cases}
 \frac{2}{{\sqrt c }}\arctan \frac{{\sqrt c }}{2},&\alpha = 0 \\
 \frac{2}{{c\ln 2}}\ln \frac{{c\ln 2 + 2}}{2},&\alpha = 1 \\
 \frac{{2(\alpha - 1)}}{{2(\alpha - 1) + c e^{- c \frac {{\alpha-2}}{{\alpha-1}}} }},&\alpha \ge 2. \\
\end{cases}
\end{equation}
The above results suggest that the relationship between the exact time cost
and the distance is linear for most values of $\alpha$. Meanwhile, we have studied
the size effect on navigation with cost constraint. Based on  Eq.(\ref{lambda}), we
 know that $\lambda$ approaches its limit much more slower when $\alpha$ gets closer
  to 2 as $n$ increases. The time cost of navigation with different network
  sizes are provided in FIG.\ref{sizeeffect}. It can be verified that it will approach
  its limit as $n$ goes to infinity, which is given by Eq.(\ref{time cost}).

In summary, we constructed a the dynamical equation for the
1-dimensional navigation with limited cost. Based on the equation,
we proved that for large networks and comparatively
small cost the optimal power-law exponent is $\alpha=2$. Our analytical
results confirm the previous simulations\cite{Hav10} well.

\textbf{Acknowledgement.}We wish to thank Prof. Shlomo Havlin for
some useful discussions and two anonymous referees for their helpful
suggestions. This work is partially supported by the
Fundamental Research Funds for the Central Universities and NSFC
under Grant No. 70771011 and 60974084. Y. Hu is supported by
Scientific Research Foundation and Excellent Ph.D Project of Beijing
Normal University.

\end{document}